\documentclass[12pt]{article}

\advance\hoffset by -7mm   
\makeatletter 
\@addtoreset{equation}{section}  
\makeatother 
 
\def\ii{\'\i} 
\def\ao{\~ao}

\def\cao{\c c\~ao}

\def\ftoday{{\sl {Le \number\day \space\ifcase\month  
\or janvier\or f\'evrier\or mars\or avril\or mai 
\or juin\or juillet\or ao\^ut\or septembre\or octobre 
\or novembre \or d\'ecembre\fi\space \number\year}}}     
\def\ptoday{{\sl {\number\day \space de\space \ifcase\month  
\or janeiro\or fevereiro\or mar{\c c}o\or abril\or maio 
\or junho\or julho\or agosto\or setembro\or outubro 
\or novembro \or dezembro\fi\space de\space \number\year}}}     
\def\gtoday{{\sl {Den \number\day. \ifcase\month  
\or Januar\or Februar\or M\"arz\or April\or Mai 
\or Juni\or Juli\or August\or September\or Oktober 
\or November \or Dezember\fi\space \number\year}}}     
\def\today{{\sl {\ifcase\month 
\or January\or February\or March\or April\or May 
\or June\or July\or August\or September\or October 
\or November \or December\fi \space\number\day,\space  
                                            \number\year}}} 
\newcommand{\journal}[4]{{\em #1~}#2\,(#3)\,#4}

\newcommand{\prl}{\journal {Phys. Rev. Lett.}}

\newcommand{\cqg}{\journal {Class. Quantum Grav.}} 
 
\newcommand{\np}{\journal {Nucl. Phys.}} 
\newcommand{\pl}{\journal {Phys. Lett.}} 
 
\newcommand{\prep}{\journal {Phys. Rep.}}

\renewcommand{\d}{\delta}          
\newcommand{\e}{\varepsilon} 
 
\newcommand{\la}{\lambda}        \newcommand{\LA}{\Lambda} 
\newcommand{\m}{\mu} 
\newcommand{\n}{\nu} 
\newcommand{\om}{\omega}          
\newcommand{\p}{\psi}               
\newcommand{\s}{\sigma}           \renewcommand{\S}{\Sigma} 
\renewcommand{\th}{\theta}          
\newcommand{\f}{{\phi}}            
\newcommand{\vf}{{\varphi}} 

\newcommand{\CC}{{\cal C}}

\newcommand{\GG}{{\cal G}}

\newcommand{\OO}{{\cal O}}

\newcommand{\es}{\\[1.7mm]}

\newcommand{\sla}{\raise.15ex\hbox{$/$}\kern -.57em}  
\newcommand{\Sla}{\raise.15ex\hbox{$/$}\kern -.70em}

\newcommand{\lp}{\left(}\newcommand{\rp}{\right)} 
 
\newcommand{\lac}{\left\{}\newcommand{\rac}{\right\}} 
 
\newcommand{\complex}{{\kern .1em {\raise .47ex 
\hbox {$\scriptscriptstyle |$}} 
    \kern -.4em {\rm C}}} 
\newcommand{\real}{{{\rm I} \kern -.19em {\rm R}}} 
\newcommand{\rational}{{\kern .1em {\raise .47ex 
\hbox{$\scripscriptstyle |$}} 
    \kern -.35em {\rm Q}}} 
\renewcommand{\natural}{{\vrule height 1.6ex width 
.05em depth 0ex \kern -.35em {\rm N}}}

\newcommand{\pa}{\partial}

\newcommand{\dfud}[2]{{\displaystyle{\frac{\delta #1}{\delta #2}}}} 
\newcommand{\dfrac}[2]{{\displaystyle{\frac{#1}{#2}}}}

\newcommand{\twiddle}{\lower.9ex\rlap{$\kern -.1em\scriptstyle\sim$}}

\newcommand{\equ}[1]{(\ref{#1})} 
\newcommand{\eq}{\begin{equation}} 
\newcommand{\eqn}[1]{\label{#1}\end{equation}} 
\newcommand{\eea}{\end{eqnarray}} 
\newcommand{\eqa}{\begin{eqnarray}} 
\newcommand{\eqan}[1]{\label{#1}\end{eqnarray}} 
\newcommand{\ba}{\begin{array}} 
\newcommand{\ea}{\end{array}} 
\newcommand{\eqac}{\begin{equation}\begin{array}{rcl}} 
\newcommand{\eqacn}[1]{\end{array}\label{#1}\end{equation}} 
 
\newcommand{\ads}{{(A)dS}}
\newcommand{\ee}{{e^1_x}}

\begin{document} 
\thispagestyle{empty}


\bigskip 
\bigskip 


\begin{center} 
{\Large {\bf Canonical Analysis of the Jackiw-Teitelboim\\
 Model in the Temporal Gauge.\es
I. The Classical Theory}}
 
\vspace{5mm}

{\large 
Clisthenis P. Constantinidis\footnote{Work supported
   in part by the Conselho Nacional
   de Desenvolvimento Cient\'{\i}fico e
   Tecnol\'{o}gico -- CNPq (Brazil).}$^{,}$\footnote{Work supported
   in part by the PRONEX project No. 35885149/2006 from FAPES -- CNPq (Brazil).}, 
Jos\'e Andr\'e Louren\c co$^{1,2,}$\footnote{Work supported in part by 
   the Funda\cao\ de Apoio \`a Ci\^encia e
   Tecnologia do Esp\ii rito Santo -- FAPES (Brazil).}, \es
Ivan Morales$^{2,3,}$\footnote{Work supported in part by the 
   Coordena\cao\ de Aperfei\c coamento de Pessoal de N\ii vel Superior -- CAPES
(Brazil).}, 
Olivier Piguet$^{1,2}$ and  
Alex Rios$^{2,3,4}$ 
}
\vspace{5mm}

\noindent $^{*}$ {\it Universidade Federal do Esp\'{\i}rito Santo 
(UFES), 
CCE, Departamento de F\'{\i}sica, Av. Fernando Ferrari, 514, 
BR-29075-910 - Vit\'oria - ES (Brasil).}
\vspace{3mm}

{\small\tt E-mails: clisthen@cce.ufes.br, 
quantumlourenco@gmail.com,
mblivan@gmail.com,
opiguet@yahoo.com,
rios\_alex@ig.com.br}
\vspace{5mm}

{\sl January 2008}
\end{center}


\vspace{3mm}

\begin{abstract}

As a preparation for its quantization in the loop formalism, 
the 2-di\-men\-sion\-al gravitation model of Jackiw and Teitelboim is analysed
in the classical canonical formalism. The dynamics is of pure constraints
as it is well-known. 
A partial gauge fixing of the
temporal type being performed, the resulting second class 
constraints are sorted out and the corresponding
Dirac bracket algebra is worked out. Dirac observables of this classical
theory are then calculated.

\end{abstract}


\section{Introduction}

The full quantization of General Relativity remains an open problem
despite the very important advances achieved during the last two
decades,
mainly within the``loop quantization'' formalism of Ashtekar,
Rovelli, Smolin and others (see, e.g., the books and 
review~\cite{ash-lewan,rovelli_book,thiemann_book}. 
It is thus still worthwhile to
investigate lower dimensional gravitation theories, where the technical
difficulties of the 4-dimensional theory are somewhat 
milder~\cite{JT}~--~\cite{berg-mey}.

The purpose of the present paper is to investigate the canonical
formulation of the model of two-dimensional gravity proposed
independently by Jackiw and Teitelboim (JT) some time ago~\cite{JT}. The
JT model contains a dilaton-type scalar field, beyond the
metric field, in order to have an action which does not reduce to 
boundary terms. Moreover a cosmological constant is introduced in order
to be able to write a non-degenerate action. The invariance of the
theory is thus de Sitter or anti-de Sitter (``\ads''). We shall start from
the ``$BF$'' formulation introduced in~\cite{isler}, which explicitly
identifies the JT model as a topological gauge
theory~\cite{BF-com-vinculos}, the gauge group being \ads.

Since most approaches to loop gravity are based -- explicitly or implicitly
-- on a partial gauge fixing of the temporal type (``temporal
gauge'')~\cite{ash-lewan,rovelli_book,thiemann_book}, we shall choose to
work with a 2-dimensional version of the temporal gauge. The focus of this paper
will be on the classical theory, the quantization being left for
future publication~\cite{paper-prepa}.

After recalling the main features of the JT model in Section \ref{JT_model}, 
we spell down
the canonical formulation of the $BF$ version of the theory in 
Section \ref{Canonical_Formalism}.
Using the quantization scheme of 
Dirac~\cite{dirac} for theories with constraints, we separate, 
in section \ref{gauge_temporal}, 
the second
class constraints originated from the temporal gauge fixing, and show that the
remaining first class constraints generate a gauge symmetry which is
equivalent -- up to field equations -- to the invariance under 
space-time diffeomorphisms. The classical Dirac observables are
shortly discussed in Section \ref{observables}, and some brief conclusions are given in
Section \ref{conclusion}. 

Part of the material of the present paper has been included 
by two of the authors~\cite{alex,ivan} in  thesis presented as a
requirement to the obtention of the Master degree.

\section{The Jackiw--Teitelboim Model}\label{JT_model}

\subsection{The Jackiw--Teitelboim Action} 

Pure gravity in 2 space-time dimensions cannot be based on the Einstein-Hilbert
action $\int d^{2}x\sqrt{-g}R$, which is a surface integral,
corresponding to an identically vanishing Einstein tensor: 
$R_{\mu \nu}-\frac{1}{2}g_{\mu \nu}R \equiv0$~\cite{JT}. A simple but
nontrivial model has been proposed long ago independently by Jackiw and
by Teitelboim~\cite{JT}. The model contains,    besides the space-time
metric $g_{\mu \nu}(x)$, a dilaton type scalar field $\p(x)$. Its action
is given by
\begin{eqnarray}
S_{\rm JT}=\frac{1}{2}\int d^{2}x \sqrt{-g}\psi (R-2k)\,.\label{2}
\end{eqnarray}
It is  invariant under space-time diffeomorphisms
and leads to the Liouville equation 
\begin{eqnarray}
R-2k = 0\,,
\label{1}	
\end{eqnarray}
and to the equation for $\p$~\cite{marc-h},
\begin{eqnarray}
\nabla_\mu\nabla_\nu\psi+kg_{\mu\nu}\psi=0\,, 
\end{eqnarray} 
where
$\nabla_\mu$ is the Levi-Civitta covariant derivative associated to the metric
$g_{\mu\nu}$. Eq. \equ{1} yields a geometry with constant curvature, the
parameter $k$ playing the role of the cosmological constant.

A canonical quantization of this model in terms of the variables $g_{\mu \nu}$
and $\p$ has been given by Henneaux~\cite{marc-h}.

\subsection{$BF$ Formulation of the Jackiw-Teitelboim
Model}\label{BF_formulation}

The model is equivalent to a $BF$ model based on the gauge group \ads, i.e. 
the 2-dimensional de Sitter or anti-de Sitter group, SO(1,2) ou 
SO(2,1), according to the sign of the cosmological constant $k$~\cite{isler,fuk-kam}. 
The \ads\ gauge connection is written as
\eq
A(x)=e^{I}(x)P_{I}+\omega(x)\Lambda\,, 
\eqn{ads-connection}
where the operators $P_{I}$ ($I=0,1$) and $\Lambda$ are the ``translation''
generators and the Lorentz boost generator, respectively, 
obeying  the \ads\ algebra\footnote{By convention the antisymmetric  tensor
$\epsilon_{IJ}$ is defined by $\epsilon_{01}=1$.
The indices $I,\,J,\,\cdots=0,1$ are lowered and raised by the ``flat'' metric 
$\eta_{IJ} = $ diag$(\s,1)$ or its inverse $\eta^{IJ}$, 
where $\s=\pm1$ for the Riemannian, resp.
Lorentzian theory.} 
\begin{eqnarray}
\left[\Lambda,P_{I}\right]=\epsilon_{I}\,^{J}P_{J} \quad, \quad
\left[P_{I},P_{J}\right]=k\epsilon_{IJ}\Lambda\,. \label{sitter-un}
\end{eqnarray}
The coefficients in \equ{ads-connection} are the zweibein and Lorentz connection forms
\eq
e^I =e^I_\m dx^\m\,,\quad \om= \om_\m dx^\m\,.
\eqn{2-bein-connection}
The space-time metric is given in terms of the zweibein by
\eq
g_{\m\n}=\eta_{IJ}e^I_\m e^J_\n\,.
\eqn{metric}
Introducing the indices $i,j,\cdots=0,1,2$ and denoting the 
generators of (A)dS as $J_i$: 
\begin{eqnarray}
\{J_{i}\}=\{J_{0},J_{1},J_{2}\}=\{P_{0},P_{1},\Lambda\} \,.
\label{geradores-ads} \end{eqnarray}
the algebra \equ{sitter-un} reads
\begin{eqnarray}
\left[J_{i},J_{i}\right]=f_{ij}\,^{k}J_{k}=k\epsilon_{ijl}k^{lk}J_{k}\,,
\label{sitter dos} \end{eqnarray} 
where the nonzero structure constants $f_{ij}\,^{k}$ 
are\footnote{The completely antisymmetric tensor $\epsilon_{ijl}$
is defined by $\epsilon_{012}=1$.}:
\[
f_{01}{}^{2}=k\,,\quad f_{12}{}^{0}=\sigma\,,\quad f_{20}{}^{1}=1\,.
\]
(A)dS possesses an invariant nondegenerate 
quadratic form  $\left\langle
J_{i},J_{j}\right\rangle = k_{ij}$, where $k_{ij}$ is the Killing metric
\begin{eqnarray}
k_{ij}=-\frac{\sigma}{2}f_{ik}\,^{l}f_{j\,l}\,^{k}\,. 
\label{killing1}\end{eqnarray} 
Explicitly:
\begin{eqnarray} (k_{ij})=\left( \begin{array}{cc}
k\eta_{IJ} & 0 \\ 0 & 1 \\ \end{array} \right)\,.
\label{killing}\end{eqnarray}
This metric and its inverse are used to lower and raise the
indices $i,j,\cdots$. Notice that a nonvanishing cosmological constant $k$ 
is necessary in order to ensure the nondegeneracy of the Killing metric.

The ``$B$-field'' of the theory is a Lie algebra valued scalar field
\eq
\phi =\phi^{i}J_{i} =: \vf^I P_I+\psi\Lambda\,.
\eqn{B-field} 
With the Yang-Mills curvature given 
by\footnote{The wedge symbol $\wedge$ for exterior products of forms is ommitted.}
\eq
F= F^{i}J_{i}\equiv F^IP_I+F^2\Lambda= dA+A\,A
=\frac{1}{2}F_{\mu\nu}^{i}dx^{\mu}dx^{\nu}J_{i}\,,
\eqn{YM-curvature}
the ``$BF$'' action reads~\cite{fuk-kam,BF-com-vinculos}:
\eq 
S_{BF}[A,\phi] = \int\left\langle \phi,F\right\rangle=
\frac{1}{2}\int d^{2}x \epsilon^{\mu\nu} k_{ij}
\phi^{i}F^{j}_{\mu\nu} =: \int dt L_{BF} 
\eqn{3}
where the Lagrangian $L_{BF}$ explicitly reads\footnote{The values of the 
space-time indices $\m,\n,\cdots$ are denoted by $t$, $x$.
The antisymmetric Levi-Civitta tensor $\epsilon^{\mu\nu}$ is defined
by $\epsilon^{tx} = +1$.}
\begin{eqnarray} 
L_{BF}=\int dx(\phi_{i}\partial_{t}A^{i}_{x}+A^{i}_{t}\,D_{x}\phi_{i})\,. 
\label{3-1}\end{eqnarray}
Notice that the curvature components,
\eq\ba{l}
F^{I} = dA^{I}+f_{jk}\,\!\!^{I}A^{j}\wedge A^{k}
=de^{I}+\omega^{I}\,\!\!_{J}\wedge e^{J}\,, \es
F^{2}= dA^{2}+\frac{1}{2}f_{jk}\,\!\!^{2}A^{j}\wedge A^{k}
=d\omega+\frac{k}{2}e^{I}\wedge e^{J} \epsilon_{IJ}\,, 
\ea\eqn{curvatures}
represent the torsion $T^I:=F^I$ and the Riemann curvature with
cosmological term added, respectively.

The action (\ref{3}), which is invariant under the \ads\ gauge transformations, 
turns out to be automatically invariant under the diffeomorphisms, on shell, as a
general result for topological theories of this 
type~\cite{BF-com-vinculos}.

The field equations are
\eq
\dfud{S_{BF}[A,\phi]}{\Phi} = 0\,,\quad\Phi = \phi_i\,,\ A^i\,,
\eqn{eqm}
where
\[
 \frac{ \delta S_{BF}[A,\phi]}{\delta\phi_{i}}=F^{i}=0\,, \quad 
\frac{\delta S_{BF}[A,\phi]}{\delta A^{i}} = D\phi_{i} \,. 
\]
In components, with the notation
\begin{equation} \begin{array}{lll}
\left(\phi_i\right)&=\left(\phi_0,\phi_1, \phi_2\right)
&=: \left(\varphi_{0},\varphi_{1},\psi\right)\,,\es
\left(A^i_x\right) &= (e^0_x,e^1_x,\omega_x)
&=: (\chi,e^1_x,\omega_x)\,,\es
\left(A^i_t\right) &= (e^0_t,e^1_t,\omega_t)
&=: (N,N^1,\omega_t)\,. 
\end{array} \end{equation} 
the functional derivatives read
\begin{eqnarray}
\frac{\delta S_{BF}[A,\phi]}{\delta\varphi_{0}}
&=&\partial_{t}\chi-\partial_{x}N-\sigma(e^{1}_{x}\omega_{t}-\omega_{x}N^{1})\,,
\nonumber\\
\frac{ \delta S_{BF}[A,\phi]}{\delta\varphi_{1}}
&=&\partial_{t}e^{1}_{x}- \partial_{x}N^{1}-\omega_{x}N+\chi N^{1}\,,	
\nonumber\\
\frac{ \delta S_{BF}[A,\phi]}{\delta \psi}
&=&\partial_{t}\omega_{x}-\partial_{x}\omega_{t}-
k(\chi N^{1}-e^{1}_{x}N)\,.\nonumber\\
\frac{\delta S_{BF}[A,\phi]}{\delta\chi}
&=&-D_{t}\varphi_0=-(\partial_t\varphi_{0}+kN^1\psi-\omega_t\varphi_1)\,,\nonumber\\
\frac{\delta S_{BF}[A,\phi]}{\delta e^{1}_{x}}
&=&-D_{t}\varphi_1=-(\partial_t\varphi_{1}+\sigma \omega_t\varphi_0-kN\psi)\,,
\label{1eq}\\
\frac{\delta S_{BF}[A,\phi]}{\delta \omega_x}&=&-D_{t}\psi
=-(\partial_t\psi+N\varphi_1-\sigma N^1\varphi_0)\,.\nonumber\\
\frac{\delta S_{BF}[A,\phi]}{\delta N}&=&D_{x}\varphi_0
=\partial_x\varphi_{0}+ke^{1}_{x}\psi-\omega_x\varphi_{1}\,,\nonumber\\
\frac{\delta S_{BF}[A,\phi]}{\delta N^1}&=&D_{x}\varphi_1
=\partial_x\varphi_{1}+\sigma \omega_x\varphi_0-k\chi\psi\,,\label{eq1}\nonumber\\
\frac{\delta S_{BF}[A,\phi]}{\delta \omega_t}&=&D_{x}\psi
=\partial_x\psi+\chi\varphi_1-\sigma \omega_x\varphi_0\,.\nonumber
\end{eqnarray}
The two equations which correspond to the variation of the
scalar fields $\vf_I$ lead to the conditions of zero torsion. Solving
them for $\om_x$ and $\om_t$ in terms of the zweibein components $e^I_\m$
shows the equivalence of the $BF$ theory with the 
Jackiw-Teitelboim theory~\cite{isler,fuk-kam}. 

\section{Canonical Formalism}\label{Canonical_Formalism}

As usual in the canonical formalism (see, e.g.,~\cite{wald}), 
we assume for the space-time a
topological structure of the form 
$\mathcal{M}=\real\times\Sigma$, where the real line $\real$ represents
``time'', and 
$\Sigma$ is a 1-dimensional manifold of 
arbitrary but fixed topology, representing ``space''. Choosing the
components $A^i_\mu(x,t)$ of the connection as the generalized
coordinates, the corresponding momenta will 
be\footnote{In the following, only the dependence on the spatial
coordinate, denoted by $x$, $y$, etc., will be written explicitly.}: 
\begin{eqnarray}
\pi_{i}^{A_{x}}(x)&=&\frac{\delta{L_{BF}}} {\delta( \partial_{t}
A^{i}_{x}(x))} =\phi_{i}\,, \end{eqnarray} \begin{eqnarray}
\pi_{i}^{A_{t}}(x)&=&\frac{\delta{L_{BF}}} {\delta( \partial_{t}
A^{i}_{t}(x))} = 0\,,
\label{momento At} \end{eqnarray} 
where $L_{BF}$ is the Lagrangian \equ{3-1}
The last equation indicates that we have a singular Lagrangian and must
appeal to Dirac's formalism~\cite{dirac,henneaux-teit}. This equation
amounts to the presence of 3 primary constraints
\eq
\pi_{i}^{A_{t}}(x) \approx 0\,,\quad i=0,1,2\,,
\eqn{prim-constr}
where the symbol $\approx$ means ``weak'' equality according to the terminology 
of Dirac: such constraints will be solved only after all
Poisson algebra calculations  have been performed. 

A Legendre transformation yields the Hamiltonian
\begin{eqnarray} 
H=-\int dx A^{i}_{t}(x)D_{x}\phi_{i}(x).
\label{hamiltoniano} \end{eqnarray}
The Poisson bracket algebra is defined by the brackets of the
generalized coordinates and their conjugate momenta. The nonvanishing
ones are
\begin{eqnarray} \left\{A^{i}_{x}(x),\phi_{j}(y)\right\}
=\delta^{i}_{j}\,\delta(x-y)= \left\{A^{i}_t(x),\pi^{A_t}_{j}(y)\right\}
\end{eqnarray} 
Consistency of the dynamics requires that the primary constraints do not
  evolve, hence must have (weakly) vanishing Poisson brackets with the
Hamiltonian: 
\begin{eqnarray}
\dot{\pi}^{A_{t}}_{i}(t,x)=\left\{\pi^{A_{t}}_{i}(t,x),H\right\}\approx
0. \end{eqnarray} 
It results from
\[
\left\{\pi^{A_{t}}_{i}(t,x),H\right\}  = \partial_{x}\phi_{i}
+f_{ij}\,^{k}A^{j}_{x}\phi_{k}=D_{x}\phi_{i}(x)\,,
\]
that we must impose the secondary constraints
\begin{eqnarray} 
\mathcal{G}_{i}(t,x):= D_{x}\phi_{i}(x)	\approx 0\,. 
\label{vs1} \end{eqnarray} 
We observe that the Hamiltonian \equ{hamiltoniano} is made of 
constraints only, which 
is expected in a generally covariant
theory~\cite{henneaux-teit, tyutin}. The fields $A^{i}_{t}$, which are
not dynamical, play the role of Lagrange multipliers.

The primary constraints \equ{prim-constr} being trivially solved, we are left with
the secondary constraints \equ{vs1}. These constraints are first class
according to Dirac's terminology, since they form a closed Poisson bracket
algebra. This algebra is best expressed in terms of the ``smeared
constraints''
\eq
\mathcal{G}(\epsilon) = \int dx \;\epsilon^{i}(x)\mathcal{G}_{i}(x)\,,
\eqn{integrated-constr} 
where $\epsilon^{i}$ are arbitrary smooth functions.
Then
\begin{eqnarray}
\left\{\mathcal{G}(\epsilon),\mathcal{G}(\eta)\right\}=
\mathcal{G}(\epsilon\times\eta)\,,
\label{al1} \end{eqnarray}
where $\left(\epsilon \times \eta
\right)^{k}=f_{ij}\,^{k}\epsilon^{i}\eta^{j}$: this is the local version
of the Lie algebra of the group \ads. These constraints generate the
\ads\ gauge transformations:
\begin{eqnarray}
\left\{\mathcal{G}(\epsilon),A^{p}_{x}(x)\right\}&=
&\partial_{x}\epsilon^{p}+f_{ji}\,^{p}A^{j}_{x}\epsilon^{i}=
D_{x}\epsilon^{p}\nonumber\\
\left\{\mathcal{G}(\epsilon),\phi_{p}(x)\right\}&=&-f_{pi}\,
^{k}\epsilon^{i}\phi_{k} = \lp\phi\times\epsilon\rp_p\,.
\end{eqnarray} 

\section{Partial Gauge Fixing}\label{gauge_temporal} 
\subsection{The Temporal Gauge}

Following an approach commonly used for the 4-dimensional theory, 
as described e.g. in the review~\cite{ash-lewan}, we introduce a partial gauge
fixing, the ``temporal gauge'', which consists in making vanish the
component $\chi$ $:=$ $e^0_x$ of the zweibein. This condition is
implemented as a new constraint,
\eq
\chi\approx 0\,,
\eqn{cond-de-gauge}
 by modifying the action (\ref{3}) as:
\begin{eqnarray} 
S&=&\int d^{2}x(\phi_{i}F^{i}+B\,\chi) = \int\int dt
dx((\partial_{t}A^{i}_{x})\phi_{i}+A^{i}_{t}\,D_{x}\phi_{i}+B\,\chi)\,,
\end{eqnarray}
where $B$ is a Lagrange multiplier field. This will introduce second
class constraints which will be treated using Dirac's 
formalism~\cite{dirac,henneaux-teit}.

The conjugate momenta  and the nonvanishing Poisson brackets are now
\begin{eqnarray}
\pi^{A_{x}}_{i}(x)&=&\phi_{i}(x)\,,\quad \pi^{A_{t}}_{i}\approx
0\,,\quad \pi^{B}\approx 0\,, 
\label{vin3}\end{eqnarray}
and
\begin{eqnarray}
\left\{A^{i}_{x}(x),\phi_{j}(y)\right\}&=&\delta^{i}_{j}\delta
(x-y)\,,\nonumber\\ \left\{B(x),\pi^{B}(y)\right\}&=&\delta
(x-y)\,,\nonumber\\
\left\{\pi^{A_{t}}_{j}(x),A^{i}_{t}(y),\right\}&=&\delta^{i}_{j}\,\delta
(x-y)\,.\nonumber 
\end{eqnarray} 
We have now four primary constraints (the last two weak equalities in
\equ{vin3}). Repeating the argument of the preceding section, we arrive at
the Hamiltonian
\begin{eqnarray} H=-\int dx(A^{i}_{t}\,D_{x}\phi_{i}+B\,\chi)\,.
\end{eqnarray}
and four secondary constraints:
\eq\ba{l}
\mathcal{G}_{i} := D_{x}\phi_{i} \approx 0\,,\quad (i=0,1,2)\es
\mathcal{G}_{3} := \chi \approx 0\,. 
\ea\eqn{four_constraints}
The Poisson brackets of these constraints read, in matrix notation and
up to constraints:
\begin{eqnarray}
\begin{array}{ll}
\left\{\mathcal{G}_{\alpha}(x),\mathcal{G}_{\beta}(y)\right\}
={\CC}_{\alpha \beta}(x,y) \!\approx\! \left( \begin{array}{cccc}
\!0 &\! 0 &\! 0 &\!\!-\partial_{x}\\ \!0 & \!0 & \!0
&\!\!-\sigma\,{\omega_x}\\ \!0 & \!0 & \!0 \!&\!\!\sigma e^{1}_{x}\\
\!-\partial_{x} & \!\sigma {\omega_x} & \!-\sigma e^{1}_{x} &\!\! 0\\
\end{array} \right)\!\delta(x\!-\!y) 
\end{array}
\label{matriz-vinculo} \end{eqnarray}
with $\alpha,\beta=0,1,2,3$. The rank of the matrix 
${\CC}_{\alpha \beta}$ is equal to 2, which means that we have two
second class constraints. In order to separate them from the first
class ones, we proceed to a redefinition
\begin{eqnarray}
\mathcal{G}'_{0}(x)&=&(e^{1}_{x})\mathcal{G}_{0}(x)
-\sigma(\partial_{x}e^{1}_{x})\mathcal{G}_{2}(x) +\sigma
e^{1}_{x}\partial_{x}\mathcal{G}_{2}(x)\,, \label{g'0}\nonumber\\
\mathcal{G}'_{1}(x)&=&e^{1}_{x}(x)\mathcal{G}_{1}(x)
+{\omega_x}(x)\,\mathcal{G}_{2}(x) \label{g'1}\,,\nonumber\\
\mathcal{G}'_{2}(x)&=&\mathcal{G}_{2}(x)\,,\nonumber\\
\mathcal{G}'_{3}(x)&=&\mathcal{G}_{3}(x)\,.
\label{redef-constr} \end{eqnarray}
The new Poisson bracket matrix is
\begin{eqnarray} 
{\CC}'_{\alpha \beta}(x,y)\approx
\left( \begin{array}{cc}	0 & 0 \\ 0 & {\CC}'_{ab}(x,y) \\
\end{array} \right)\delta(x-y)\,, 
\end{eqnarray}
with the $2\times2$ submatrix $(a,b=2,3)$ given by: 
\begin{eqnarray} 
{\CC}'_{ab}(x,y)= \left( \begin{array}{cc}	0 & \sigma
e^{1}_{x} \\ -\sigma e^{1}_{x} & 0 \\ \end{array} \right)\delta(x-y)\,.
\end{eqnarray}
The latter has an inverse,
\eq
{\CC}'^{ab}(x,y) := ({\CC}'_{ab}(x,y))^{-1}=\left(
\begin{array}{cc}	0 & -\sigma/ e^{1}_{x} \\ \sigma /e^{1}_{x} & 0
\\ \end{array} \right)\delta(x-y)\,,
\eqn{inverse-C'}
in the convolution sense, i.e.:
\[
\int dz
{\CC}'^{ab}(x,z)\,{\CC}'_{bc}(z,y)
=\delta^{a}_{c}\delta(x-y)\,.
\] 
We conclude that the constraints $\mathcal{G}'_{0}$ and  $\mathcal{G}'_{1}$
are first class, whereas $\mathcal{G}'_{2}$ and
$\mathcal{G}'_{3}$ are second class.

\subsection{Dirac Brackets}

In order to take care of the second class constraints, continuing to
follow Dirac, we define the Dirac bracket between two fields $A$ and $B$ 
-- local functionals of the fields $e^{1}_{x}$, ${\omega_x}$, $\varphi_{0}$, 
$ \varphi_{1}$, $\psi$ as:
\begin{eqnarray}
&&\left\{A(x),B(y)\right\}_{D}=\left\{A(x),B(y)\right\}
{}\nonumber\\&&{} 
- \int
d^{3}z_{1}d^{3}z_{2}\left\{A(x),
\mathcal{G}'_{a}(z^{1})\right\}{\CC}^{'ab}(z_{1},
z_{2})\left\{\mathcal{G}'_{b}(z_{2}),B(y)\right\}\,,	
\end{eqnarray}
where ${\CC}^{'ab}$ is the matrix \equ{inverse-C'}.
For 
$A$ and $B$ $=$ $e^{1}_{x}$, ${\omega_x}$, $\varphi_{1}$, $\psi$ 
(but not $\varphi_{0}$), the result is simply:
\begin{eqnarray} \left\{A(x),B(y)\right\}_{D}=
\left\{A(x),B(y)\right\}\,.
\end{eqnarray}
Moreover, the Dirac bracket of any field $A$ with a second class constraint is vanishing:
\begin{eqnarray}
\left\{A(x),\mathcal{G}'_{a}(y)\right\}_{D}\!\!\!&=&\!\!\! 0 \,,\quad
a=2,3\,,\ \forall A(x) \,, 
\label{fuertemente}\end{eqnarray}
which allows to impose the second class constraints as strong equalities:
\begin{eqnarray} 
\mathcal{G}'_{2}=0\,,\quad
\mathcal{G}'_{3}=\chi=0\,. \label{fuertemente-dos} 
\end{eqnarray} 
The second equality is just the temporal gauge condition, and the
first one allows to express $\vf_0$ as a function of the other basic 
fields
\begin{eqnarray}
\varphi_{0}=\sigma\frac{\partial_{x}\psi}{e^{1}_{x}}\,. 
\label{psi-cero}\end{eqnarray}
With this, the first class constraints $\GG_0'$ and $\GG_1'$ become
\begin{eqnarray} 
\mathcal{G}'_{0}(x)&=&(e^{1}_{x})\mathcal{G}_{0}(x)
=\sigma e^{1}_{x}\partial_x\!\!\left(\frac{\partial_x\psi}{e^1_x}\right)
+ k(e^1_x )^2\psi -e^1_x{\omega_x}\varphi_1\,, \label{g0}\\[0,2 cm] 
\mathcal{G}'_{1}(x)&=&e^{1}_{x}\mathcal{G}_{1}(x) =
e^{1}_{x}\partial_{x}\varphi_{1}+{\omega_x}\partial_{x}\psi\,. \label{g1}
\end{eqnarray}
The Dirac bracket algebra of these contraints is closed:
\begin{eqnarray}
\left\{\mathcal{G}'_{0}(\epsilon),\mathcal{G}'_{0}(\eta)\right\}_{D}\!
&=&\sigma\,\mathcal{G}'_{1}([\epsilon,\eta])\,,\nonumber\\
\left\{\mathcal{G}'_{0}(\epsilon),\mathcal{G}'_{1}(\eta)\right\}_{D}\!
&=&-\,\mathcal{G}'_{0}([\epsilon,\eta])\,,
\label{alg3}\\
\left\{\mathcal{G}'_{1}(\epsilon),\mathcal{G}'_{1}(\eta)\right\}_{D}\!
&=&-\mathcal{G}'_{1}([\epsilon,\eta])\,,
\nonumber
\end{eqnarray}
where 
$[\epsilon,\eta]\,=
(\epsilon\partial_{x}\eta-\eta\partial_{x}\epsilon)\,$, 
which confirms
that $\mathcal{G}'_{0}$ and $\mathcal{G}'_{1}$ are first class.

\subsection{Gauge Symmetry and Invariance Under the Diffeomorphisms} 

The independent dynamical variables are now the fields $e^1_x$, $\om_x$,
$\vf_1$ and $\p$. Their nonvanishing Dirac brackets are
\eq
\lac e^1_x(x),\vf_1(y)\rac_{\rm D} = \d(x-y) = \lac \om_x(x),\p(y)\rac_{\rm D} \,.
\eqn{basic-dirac-brackets}

The constraints $\mathcal{G}'_{0}$ and $\mathcal{G}'_{1}$ generate
the following gauge transformations, which are  symmetries of the theory:
\eq\ba{l}
\lac \GG_0'(\e),\vf_1(y) \rac_{\rm D} = \s\dfrac{1}{\ee}\pa_x\lp\e\pa_x\p\rp 
+ 2k\e\ee\p - \e\om_x\vf_1 \,,\\[4mm]
\lac \GG_0'(\e),\p(y) \rac_{\rm D} = -\e\ee\vf_1 \,,\\[4mm]
\lac \GG_0'(\e),e^1_x(y) \rac_{\rm D} = \e\ee\om_x \,,\es
\lac \GG_0'(\e),\om_x(y) \rac_{\rm D} 
= -\s\pa_x\lp \dfrac{1}{\ee}\pa_x\lp\e\ee\rp \rp - k\e(\ee)^2 \,,
\ea\eqn{g-tr-G_0}
and, for $\mathcal{G}'_{1}$:
\begin{equation} \begin{array}{ll}
\left\{\mathcal{G}'_{1}(\eta),\varphi_{1}(y)\right\}_{D}=
\eta(y)\partial_{y}\varphi_{1}(y)\,,\\[2,5 mm]
\left\{\mathcal{G}'_{1}(\eta),\psi(y)\right\}_{D}\,\,=
\eta(y)\partial_{y}\psi(y)\,,\\[2,5mm]
\left\{\mathcal{G}'_{1}(\eta),e^{1}_{x}(y)\right\}_{D}\,=
\partial_{y}(\eta(y)\,e^{1}_{x}(y))\,,\\[2,5mm]
\left\{\mathcal{G}'_{1}(\eta),{\omega_x}(y)\right\}_{D}\,\,=
\partial_{y}(\eta(y)\,{\omega_x}(y))\,.
\label{gauge g1} \end{array} \end{equation}
These infinitesimal gauge transformations can be rewritten as\footnote{For the
transformations generated by $\GG_0'$, the field equations \equ{1eq} are
used and some heavy algebraic manipulations are necesssary.}
\begin{eqnarray}
\left\{\mathcal{G}^{'}_{0}(\epsilon),\varphi_{1}(y)\right\}_{D}
\!&=
&\!\frac{\epsilon}{e^{1}_{x}}\mathcal{G}^{'}_{0}(y)+\frac{\xi^{x}}{e^{1}
_{x}}\mathcal{G}^{'}_{1}(y)+\xi^{t}\frac{\delta S_{BF}[A,\phi]}{\delta
e^{1}_{x}}{}\nonumber\\&&{}
-\frac{\sigma}{e^{1}_{x}}\lambda\,\partial_{y}\psi
+\mathcal{L}_{(\xi^{t},-\xi^{x})}\varphi_{1}(y)\label{g0 p1}\,, \\[0,25
cm] 
\left\{\mathcal{G}^{'}_{0}(\epsilon),\psi(y)\right\}_{D}
\!&=&\!\xi^{t}\frac{\delta S_{BF}[A,\phi]}{\delta {\omega_x}}
+\mathcal{L}_{(\xi^{t},-\xi^{x})}\psi(y)\label{g0 p2}\,, \\[0,25 cm]
\left\{\mathcal{G}^{'}_{0}(\epsilon),e^{1}_{x}(y)\right\}_{D}
\!&=&\!-\xi^{t}\frac{\delta S_{BF}[A,\phi]}{\delta\varphi_{1}}
+\mathcal{L}_{(\xi^{t},-\xi^{x})}e^{1}_{x}(y)\label{g0 p3}\,, \\[0,25
cm] 
\left\{\mathcal{G}^{'}_{0}(\epsilon),{\omega_x}(y)\right\}_{D}\!&=&\!-
\xi^{t}\frac{\delta S_{BF}[A,\phi]}{\delta\psi}\!+\!
\sigma\partial_{y}\lambda\!+\!\mathcal{L}_{(\xi^{t},-
\xi^{x})}{\omega_x}(y)\label{g0 p4}, 
\end{eqnarray} 
where 
\begin{eqnarray}
\lambda=\frac{\epsilon}{N}\partial_yN-
\frac{\epsilon}{e^1_x}\partial_ye^1_x-\partial_y\epsilon\,,\quad\xi^{t}=
\frac{\epsilon e^1_x}{N}\,,\quad\xi^{x}=\frac{\epsilon N^1}{N}\,. 
\end{eqnarray} 
and
\begin{equation} \begin{array}{ll}
\left\{\mathcal{G}'_{1}(\eta),\varphi_{1}(y)\right\}_{D}=
\mathcal{L}_{(0\,,\,
\eta)}\varphi_{1}(y)\\[2,5 mm]
\left\{\mathcal{G}'_{1}(\eta),\psi(y)\right\}_{D}\,\,=
\mathcal{L}_{(0\,,\,\eta)}\psi_{1}(y)\\[2,5mm]
\left\{\mathcal{G}'_{1}(\eta),e^{1}_{x}(y)\right\}_{D}\,=
\mathcal{L}_{(0\,,\,
\eta)}e^{1}_{x}(y)\\[2,5mm]
\left\{\mathcal{G}'_{1}(\eta),{\omega_x}(y)\right\}_{D}\,\,=
\mathcal{L}_{(0\,,\,\eta)}{\omega_x}(y)
\label{gauge g1'} \end{array} \end{equation}
In the expressions (\ref{g0 p1}--\ref{g0 p4}, \ref{gauge g1'}), the symbol
$\mathcal{L}_{(v^{t},v^{x})}$ represents the Lie derivative in the direction of 
the vector $(v^{t},v^{x})$, which generates the time and space diffeomorphisms. 
The interpretation of this result is as follows. 
The time gauge condition (\ref{cond-de-gauge}), 
which breaks gauge invariance, leaves two residual symmetries unbroken. The first one is
that of time diffeomorphisms, generated by 
$\mathcal G'_0$, up to  constraints, up to field equations 
(``on-shell realization''), and up to a compensating local 
Lorentz transfortmation of parameter $\la$ which takes care of the time
gauge condition. The second unbroken invariance is that of space
diffeomorphisms, generated by $\mathcal G'_1$.

The definition of $\mathcal{G}'_{0}$ and $\mathcal{G}'_{1}$ in
\equ{redef-constr} has been   chosen in order to  be scalar
densities of weight 1. This indeed ensures that they  form a Lie algebra 
(\ref{alg3}) which is closed -- in contrast with gravity in 
higher dimensions where the algebra closes with field dependent structure 
``constants''~\cite{rovelli_book,ash-lewan,thiemann_book}. Such a
feature is characteristic of 2-dimensional theories with general covariance, 
such as the bosonic string in the approach 
of~\cite{thiemann-string}.

A new redefinition
\begin{eqnarray}
\mathcal{C_+}&=&\frac{\sqrt{-\sigma}}{2}\,\mathcal{G}'_{0}-\frac{1}{2}\,
\mathcal{G}'_{1}\,,\\
\mathcal{C_-}&=&-\frac{\sqrt{-\sigma}}{2}\,\mathcal{G}'_{0}-\frac{1}{2}\,
\mathcal{G}'_{1}\,, 
\end{eqnarray} 
leads to the algebra
\begin{eqnarray}
\left\{{\CC}_+(\epsilon),{\CC}_+(\eta)\right\}_{D}
\!&=
&{\CC}_+([\epsilon,\eta])\,,\nonumber\\
\left\{{\CC}_-(\epsilon),{\CC}_-(\eta)\right\}_{D}
\!&=&{\CC}_-([\epsilon,\eta])\,,\\
\left\{{\CC}_+(\epsilon),{\CC}_-(\eta)\right\}_{D}\!&=&0\,.
\nonumber 
\end{eqnarray}
which shows a factorization in two classical Virasoro algebras.

To complete this section, let us write the final Hamiltonian
\begin{eqnarray}
H_{F}\!\!\!&=&\!\!\!-
\!\int\!\!dy\left(\zeta^0(y)\mathcal{G}'_{0}(y)+\zeta^1(y)\mathcal{G}'_{
1}(y)\right)\,,\label{ht1} 
\end{eqnarray} 
where $\zeta^0$ e $\zeta^1$ are scalar densities of weight $-1$ 
in 1-dimensional space.
The equations of the dynamical
fields generated by this Hamiltonian are
\begin{eqnarray}
&&\partial_{t}e^1_x(x) =  \left\{e^1_x(x),H_{F}\right\}_{D}=
\zeta^0(x)e^1_x(x)\omega_x(x)+\partial_x(\zeta^1(x)e^1_x(x))\,,
\nonumber\\ %
&&\partial_{t}\omega_x(x) = \left\{\omega_x(x),H_{F}\right\}_{D}=
-
\sigma\partial_x(\partial_x\zeta^0(x)+
\frac{\zeta^0(x)}{e^1_x(x)}\partial_xe^1_x(x)) \nonumber \\
&&\qquad\qquad\qquad - 
k\zeta^0(x)(e^1_x(x))^2+\partial_x(\zeta^1(x)\omega_x(x))\,,\nonumber\\
&&\partial_{t}\varphi_{1}(x) = \left\{\varphi_{1}(x),H_{F}\right\}_{D}=
\sigma\partial_x(\zeta^0e^1_x)\frac{\partial_x\psi}{(e^1_x)^2}
+2k\zeta^0e^1_x\psi -
\zeta^0\omega_x\varphi_1+\zeta^1\partial_x\varphi_1\,,\nonumber\\ %
&&\partial_{t}\psi(x)\ = \left\{\psi(x),H_{F}\right\}_{D}=
-
\zeta^0e^1_x\varphi_1+\zeta^1\partial_x\psi\,.\nonumber 
\end{eqnarray} 
They are equivalent, modulo the constraints, to
the field equations \equ{1eq} for the fields $\ee$, $\om_x$, $\vf_1$
and $\p$.

\section{Observables}\label{observables}
\subsection{In the $BF$ Formalism}

Classical observables are gauge invariant functions in phase space. 
In Dirac's formalism, this means that they are functions $\OO$ which
have vanishing Dirac bracket with the constraints (\ref{g0}-\ref{g1}):
\begin{eqnarray}
\left\{\mathcal{O},\mathcal{G}_m\right\}_{\rm D}\approx0, \quad m=0,1\,. 
\label{op} \end{eqnarray}
We shall consider the space manifold $\S$ to be compact, homeomorphic to
the circle $S^1$. The coordinate $x$ will be denoted by $\th$, with range
$(0,2\pi)$. The nonvanishing Dirac brackets of the basic fields read
\begin{eqnarray}
\left\{e^1_x(\theta),\varphi_1(\theta')\right\}_{\rm D}=\delta(\theta-\theta')=
\left\{
\omega_x(\theta),\psi(\theta')\right\}_{\rm D}
\nonumber \end{eqnarray} 
The two independent observables present in the theory, denoted by $T$
and $L$, are defined, prior to the time gauge fixing, 
by\footnote{They were calculated by the authors 
of~\cite{livine-perez-rovelli} in the case of the compact gauge group SU(2) 
-- corresponding to \ads\ with $\s=k=1$.}
\begin{eqnarray}
	T&=&\mbox{Tr}\,Pe^{\oint_{s}A}=\mbox{Tr}\,Pe^{\oint_{s}J_iA^i}
	=\mbox{Tr}\left(\sum^{\infty}_{n=0}\frac{1}{n!}P\oint_{s_1}
\!\!\!\!A\oint_{s_2}\!\!\!\!A\cdots \oint_{s_{n-1}}\!\!\!\!A\right)
\label{olw}\,\\	
	L&=&\left\langle \phi(\theta),\phi(\theta)\right\rangle
=k^{ij}\phi_i(\theta)\phi_j(\theta)
\label{oa} \,
\end{eqnarray}
where $A$ is the \ads\ connection and $\f$ the scalar field in the
adjoint representations as defined in Subsection \ref{BF_formulation}.
$T$, defined by (\ref{olw}) is known as a Wilson loop, where 
$P$ denotes the path ordering in the $\theta$ coordinate,  and $J_i$ ($i=0,1,2$) 
are the generators of \ads. The observable $L$ defined by \equ{oa} is actually
global, too, since it is independent of $\th$ as a consequence of
the field equations.

For explicit calculations in terms of the component fields $e^I_\m$, etc.,
defined by \equ{ads-connection} and \equ{B-field}, 
it is useful to take the generators  $J_i$ in the fundamental representation as
\begin{eqnarray}
 J_0 = P_0 =-\frac{i}{2}\sqrt{k}\,\tau_3\,,\quad
 J_1 = P_1 = -\frac{i}{2}\sqrt{\sigma k}\,\tau_1\,,\quad
 J_2 = \LA = -\frac{i}{2}\sqrt{\sigma}\,\tau_2\,,\nonumber 
\end{eqnarray} 
where $\tau_1, \tau_2$ and $\tau_3$ are the Pauli matrices, the
Killing form $<\ ,\ >$ being represented by the trace.
Some useful formulae are 
\begin{eqnarray}
	J_iJ_J&=& \frac{1}{2}f_{ij}\,\!\!^{k}J_{k}-
\frac{\sigma}{4}k_{ij}\,,
	\label{rfg}
\end{eqnarray}
\eq
	\mbox{Tr}(J_iJ_j)= -\frac{\sigma}{2}k_{ij}\,,\quad
\mbox{Tr}(J_{j_1}J_{j_2}J_{j_3})
=-\frac{\sigma}{4}f_{j_{1}j_{2}}\,\!\!^{k}k_{kj_3}\,,\,\,\, etc.\
%
\eqn{trilinear}

\subsection{In the Time Gauge Formalism}

Let us now compute $T$ and $L$ for the time gauge fixed theory and check
that the resulting expressions have vanishing Dirac bracket with the
constraints $\GG_0$ and $\GG_1$. The calculations for $\GG_0$ will be
performed to the first nontrivial order of the expansion \equ{olw}.

Using explicitly the time gauge condition 
and the expression of
$\vf_0$ given from the second class constraints (see eqs.
(\ref{fuertemente-dos},\ref{psi-cero})), with the help of
(\ref{rfg},\ref{trilinear}),
we can rewrite \equ{olw} as 
\begin{eqnarray}
T&=&\mbox{Tr}(1+\oint_{s}\!\!A+\frac{1}{2!}
P\oint_{s_1}\!\!\!\!A\oint_{s_2}\!\!\!\!A+\cdots)\nonumber\\
&=&2-\frac{\sigma}{2}\int^{2\pi}_{0}\!\!\!d\theta_1
\int^{\theta_1}_{0}\!\!\!d\theta_2(e^1_x(\theta_1)e^1_x(\theta_2)k
+\omega_x(\theta_1)\omega_x(\theta_2))+\textsl{O(4)}\,,
\end{eqnarray}
where $O(4)$ means up to terms of order 4 in the basic fields.
One can then check, up to this order, that $T$ is an observable:
\begin{eqnarray}
\left\{\mathcal{G}'_0(y),T\right\}_D\approx\left\{\mathcal{G}'_1(y),T\right\}_D\approx0+\textsl{O(4)}\,.
\label{obT}
\end{eqnarray}
For the quantity $L$ given by \equ{oa}, we obtain
\begin{eqnarray}
	L&=&k^{ij}\phi_i\phi_j=\frac{\sigma}{k}(\varphi_0)^2
+\frac{1}{k}(\varphi_1)^2+(\psi)^2\,,
\end{eqnarray}
with $\varphi_0=\sigma\partial\psi/e^1_x$. It is easy to check that $L$
has weakly vanishing Dirac brackets with the constraints: 
\begin{eqnarray}
\left\{\mathcal{G}'_0(\epsilon),L\right\}_D
&=&-2\frac{\epsilon\varphi_0}{ke^1_x}\mathcal{G}'_1(x)
+2\frac{\epsilon\varphi_1}{ke^1_x}\mathcal{G}'_0(x)\approx0\,,\label{oa0}\\
\left\{\mathcal{G}'_1(\epsilon),L\right\}_D
&=&2\frac{\sigma\epsilon\varphi_0}{ke^1_x}\mathcal{G}'_0(x)
+2\frac{\epsilon\varphi_1}{ke^1_x}\mathcal{G}'_1(x)
\approx0\,.
\label{oa1}
\end{eqnarray}
Hence $L$ defines an observable, too.

\section{Conclusion}\label{conclusion}

The canonical construction of the classical theory in the time gauge has been 
completed in the Dirac formalism,  including the discussion of the observables.

This represents a first step towards the construction of the
corresponding quantum theory using the loop quantization
techniques~\cite{CLPS,paper-prepa}. 

\noindent
{\bf Acknowledgments.} We thank Alejandro Perez for very useful discussions. 


\begin{thebibliography}{99} 

\bibitem{ash-lewan} A. Ashtekar and J. Lewandowski,  ``Background independent quantum gravity: 
A Status report'', \cqg{21}{2004}R53, e-Print Archive: gr-qc/0404018.

\bibitem{rovelli_book} C. Rovelli, ``Quantum Gravity'', Cambridge Monography on Math.
    Physics (2004).

\bibitem{thiemann_book} T. Thiemann,
    ``Modern Canonical Quantum General Relativity", 
Cambridge Monographs on Mathematical Physics (2007).

\bibitem{JT} R. Jackiw, in ``Quantum Theory of Gravity'', edited by S. Christensen
(Hilger, Bristol,1984);\\
C. Teitelboim, \pl{B126}{1983}{41};\\
C. Teitelboim  in ``Quantum Theory of Gravity'', edited by S. Christensen
(Hilger, Bristol,1984).

\bibitem{marc-h} Marc Henneaux, ``Quantum Gravity in
Two-Dimension: Exact Solution of Jackiw Model'', Phys. Rev. Lett. 54
(1985) 959. 

\bibitem{fuk-kam} T. Fukuyama and K. Kamimura,
``Gauge Theory of Two-Dimensional Gravities'', Phy. Lett. 160B (1985)
259. 

\bibitem{isler} K. Isler and C.A.
Trugenberger, ``Gauge Theory of Two-Dimensional Quantum Gravity'', 
 \prl{63}{1989}{834}.

\bibitem{kum-lie-vas1} W. Kummer, H. Liebl, and D. V. Vassilevich, 
``Exact path integral quantization 
of generic 2-d dilaton gravity'', \np{B493}{1997}{491}.

\bibitem{kum-lie-vas2} W. Kummer, H. Liebl, and D. V. Vassilevich, 
``Integrating geometry in general 2d dilaton gravity
with matter'', \np{B544}{1999}{403}.

\bibitem{grum-kum-vas} D. Grumiller, W. Kummer and D.V. Vassilevich,
\prep{369}{2002}{327}

\bibitem{livine-perez-rovelli} E.R. Livine, Alejandro Perez, C. Rovelli,
``2D manifold-independent spinfoam theory'', 
\cqg{20}{2003}{4425},  [arXiv:gr-qc/0102051].

\bibitem{berg-mey} L. Bergamin and R. Meyer, 
``Two-Dimensional Quantum Gravity with Boundary'', arXiv:0711.3595 [hep-th].

\bibitem{BF-com-vinculos} D. Birmingham, M. Blau, M.
Rakowski and G.T. Thompson, ``Topological Field Theory'', Phys.Rept.
209:129-340,1991. M. Blau and G. Thompson, ``Topological Gauge Theories
of Antisymetric Tensor Field'', Ann. Phys. 205(1991) 130-172.

\bibitem{paper-prepa} C.P. Constantinidis J.A. Louren\c co
and O. Piguet,  
work in progress.

\bibitem{dirac} P.A.M. Dirac, ``Lectures on Quantum Mechanics'', Belfer
Graduate School of Science, Yeshiva University, 1964. 

\bibitem{alex} Alex Rios Costa,
``Uma Revis\ao\ da Gravita\cao\ Bidimensional do Ponto de Vista da 
Gravita\cao\ Qu\^antica de Loops'', Master Degree thesis, Universidade
Federal do Esp\ii rito Santo, Brazil (2007).

\bibitem{ivan} Luis Ivan Morales Bautista, ``Formalismo Hamiltoniano
do Modelo de Jackiw-Teitelboim
no Calibre Temporal'', Master Degree thesis, Universidade
Federal do Esp\ii rito Santo, Brazil (2007).

\bibitem{wald} R.M. Wald, ``General Relativity'', The University of
Chicago Press (Chicago and London, 1984).

\bibitem{henneaux-teit} M. Henneaux and C. Teitelboim, ``Quantization of
Gauge Systems'', Princeton University Press, 1991. 

\bibitem{tyutin} D.M.
Gitman and I.V. Tyutin, ``Quantization of Fields with Constraints'',
Springer-Verlang Series in Nuclear and Particle Physics, Berlin
Heidelberg, 1990. 


\bibitem{CLPS} C.P. Constantinidis, J.A. Louren\c co,  
O. Piguet and W. Spalenza, ``Quantiza\c c\~ao da Gravidade em 
Duas Dimens\~oes  via o Formalismo de La\c cos'', 
poster presented at the ``XXVII Encontro Nacional de F\ii sica de
Part\ii culas e Campos'',  \'Aguas de Lind\'oia, SP (2006).

\bibitem{thiemann-string} T. Thiemann, ``The LQG string: 
Loop quantum gravity quantization of string theory I: Flat target space'',
\cqg{23}{2006}{1923}, arXiv: hep-th/0401172.






\end{thebibliography}
\end{document}